\documentclass[iop,apj]{emulateapj}
\slugcomment{Received 2020 June 16; revised 2020 August 29; accepted 2020 September 15; published 2020 November 6}
\shorttitle{A Cold Quasar at $\lowercase{\emph{z}} \sim 0.405$ Observed with SOFIA HAWC+}
\shortauthors{Cooke et al. 2020}

\usepackage{natbib}
\usepackage{graphicx} 
\usepackage{textcomp}
\usepackage{comment}
\usepackage{gensymb}
\usepackage{epsfig}%
\usepackage{setspace}
\usepackage{verbatim}
\usepackage[outdir=./]{epstopdf}
\usepackage{indentfirst}
\usepackage[colorlinks=true,citecolor=blue,linkcolor=blue]{hyperref}
\usepackage{amssymb}
\usepackage{amsmath}
\usepackage{amsfonts}%
\usepackage{fancyhdr}
\usepackage[T1]{fontenc}
\usepackage[utf8]{inputenc} 

\begin{document}

\title{Dying of the Light: An X-ray Fading Cold Quasar at $\lowercase{\emph{z}} \sim 0.405$}

\author{
Kevin C. Cooke$^{1,\dagger}$, 
Allison Kirkpatrick$^{1}$, 
Michael Estrada$^{1}$, 
Hugo Messias$^{2,3}$, 
Alessandro Peca$^{4}$, 
Nico Cappelluti$^{4}$,
Tonima Tasnim Ananna$^{5,6,7}$,
Jason Brewster$^{8}$,
Eilat Glikman$^{9}$,
Stephanie LaMassa$^{10}$,
T. K. Daisy Leung$^{11,12}$,
Jonathan R. Trump$^{13}$,
Tracey Jane Turner$^{8,14}$,
C. Megan Urry$^{5,6}$
}
\affil{$^1$Department of Physics \& Astronomy, University of Kansas,
Lawrence, KS 66045, USA\\
$^2$Joint ALMA Observatory, Alonso de Cordova 3107, Vitacura 763-0355, Santiago, Chile\\
$^3$European Southern Observatory, Alonso de Cordova 3107, Vitacura, Casilla 19001, Santiago de Chile, Chile\\
$^4$Physics Department, University of Miami, Coral Gables, FL 33155, USA\\
$^5$Yale Center for Astronomy \& Astrophysics, New Haven, CT 06520, USA\\
$^6$Department of Physics, Yale University, PO BOX 201820,New Haven, CT 06520, USA\\
$^{7}$Department of Physics \& Astronomy, Dartmouth College, 6127 Wilder Laboratory, Hanover, NH 03755, USA\\
$^8$Department of Physics, University of Maryland Baltimore County, Baltimore, MD 21250, USA\\
$^9$Department of Physics, Middlebury College, Middlebury, VT 05753, USA\\
$^{10}$Space Telescope Science Institute, 3700 San Martin Drive, Baltimore, MD 21218, USA\\
$^{11}$Department of Astronomy, Space Sciences Building, Cornell University, Ithaca, NY 14853, USA\\
$^{12}$Center for Computational Astrophysics, Flatiron Institute, 162 Fifth Avenue, New York, NY 10010, USA\\
$^{13}$Department of Physics, University of Connecticut, Storrs, CT 06269, USA\\
$^{14}$Center for Space Science and Technology, University of Maryland Baltimore County, 1000 Hilltop Circle, Baltimore, MD 21250, USA
}
\email{$^\dagger$ Corresp. Author: kcooke@ku.edu}
\begin{abstract}

Cold quasars are a rare subpopulation observed to host unobscured, X-ray luminous active galactic nuclei (AGN) while also retaining a cold gas supply fueling high star formation rates.  These objects are interpreted as AGN early in their evolution. We present new SOFIA HAWC+ far-infrared observations, FUV-FIR photometry, and optical spectroscopy to characterize the accretion and star formation behavior in a cold quasar at $z \sim 0.405$ (CQ 4479).  CQ 4479 is a starburst galaxy with a predominantly young stellar population and a high gas mass fraction of $\sim50-70\%$.  The AGN component has yet to become the dominant component of the FIR emission.  We also find AGN bolometric luminosity that varies as a function of observation method and AGN region probed. Finally, we identify a candidate outflow feature corroborating the hypothesis that cold quasars have energetic feedback.  This object presents an intriguing look into the early stages of AGN feedback and probes the rare phase where an AGN and cold gaseous component coexist.
\end{abstract}




\section{Introduction}\label{sec:intro}
Galaxies observed in the local universe result from the interplay between a given galaxy’s stellar, gas, dust, and supermassive black hole (SMBH) components. In the most massive galaxies ($>10^{11}$ M$_{\odot}$), the current paradigm evokes a gas-rich, highly star-forming past that is ended via a feedback mechanism induced by an active phase of accretion onto the SMBH \citep{Henriques:2015aa,Amarantidis:2019aa}. This so-called active galactic nucleus (AGN) phase results in the injection of significant levels of radiative and mechanical energy into the interstellar medium (ISM) of its host, effectively quenching star formation \citep{Benson:2003aa}. The transition period between star-forming and quenched galaxies, and the role of the AGN in this transition, remains a poorly understood phase of galaxy evolution \citep[][and references therein]{Heckman:2014aa}. 

Characterized as unobscured X-ray sources \citep[L$_{X}$ $>$ 10$^{44}$ erg s$^{-1}$, M$_{B}$ $<$ -23,][]{Kirkpatrick:2020aa} with a cold dust component (S$_{250\;\micron}$ $>$ 30 mJy), cold quasars represent a short-lived phase where AGN and starburst coexist.  Star-bursting quasars have been previously observed, albeit at high $z$ \citep[$z > 5$;][]{Walter:2009aa,Leipski:2014aa,Decarli:2018aa}.  This has made follow-up observations difficult, especially toward the goal of constraining host galaxy properties. Cold quasars present a complementary opportunity to investigate FIR detected quasars at low redshift.  A key result from cold quasars is their enhanced WISE MIR band emission in comparison to other unobscured quasars.  \citet{Kirkpatrick:2020aa} found 72$\%$ of WISE-detected cold quasars hosted W3$_{Vega}$ $<$ 11.5, in comparison to 19$\%$ of a redshift-matched unobscured quasar population also from Stripe82X.  These MIR features point toward several potential explanations: an optically thin obscuring torus, a clumpy torus or large-scale obscuring medium, and/or contamination from the concurrent star formation episode.  To better understand the origin and structure of the central engine, we obtain new SOFIA observations to follow up on new cold quasar candidates with L$_{\rm X}$ and M$_B$ within a factor of 3 of the original classification criteria to more accurately discern the origin of the dust emission in these rare objects.

We present a case study the of the cold quasar SDSS J014040.71+001758.1 using data from the X-ray to the far-IR, including new observations using SOFIA HAWC+. These observations are used to decompose the stellar, dust, and AGN components through spectral energy distribution (SED) fitting.  We discuss the target and observations in Section \ref{sec:inst}.  The multi-wavelength emission and SED fitting process is described in Section \ref{sec:methods}, with a discussion on these results in Section \ref{sec:results}.  We summarize our conclusions in Section \ref{sec:conc}.  In this work, we assume a standard cosmology with $H_{0}$ = 70 km s$^{-1}$ Mpc$^{-1}$, $\Omega_M$ = 0.3, and $\Omega_{\Lambda}$ = 0.7 \citep{PlanckCollaboration:2016aa}.  All magnitudes are reported in the AB magnitude system unless specified otherwise \citep{Oke:1983aa}.


\section{Object and Multiwavelength Data}\label{sec:inst}


\subsection{Target Information}\label{sec:targetinfo}
SDSS J014040.71+001758.1 (hereafter CQ 4479) is a cold quasar located at R.A.J2000 of 1h 40m 40.71s  and Decl.J2000 of 0$\degree$ 17$\arcmin$ 58.17$\arcsec$  with a spectroscopic redshift of 0.40500$\pm$0.00003 from Sloan Digital Sky Survey (SDSS) Data Release 16 (DR16) \citep{Albareti:2017aa}.   CQ 4479 was originally observed as part of the Stripe 82X survey, a 31.3 deg$^2$ noncontiguous region \citep{LaMassa:2013aa,LaMassa:2013ab,LaMassa:2016aa} of the original Stripe 82 field that includes a total of 15.6 deg$^2$ of overlapping XMM-Newton and Herschel/SPIRE coverage as part of the Accretion History of AGN survey (AHA\footnote{http://project.ifa.hawaii.edu/aha/}; PI, M. Urry; see \citealt{LaMassa:2016aa} for full survey details). 

CQ 4479 was not examined in the original \citet{Kirkpatrick:2020aa} cold quasar sample due to its k-corrected X-ray luminosity ($\log_{10} L_{\rm 0.5-10keV}=43.84$), which is below the original classification threshold of $\log_{10} L_{\rm 0.5-10keV}=44.0$. In this work, we selected CQ 4479 for follow-up because of its combination of Herschel detection and classification as an optically classified broadline quasar, as well as a secure narrow-line quasar \citep{Kewley:2006aa}. We discuss this classification in Section \ref{sec:optlines}.


\subsection{X-Ray}
The source was observed by XMM-Newton in the Stripe 82X observational campaign \citep{LaMassa:2013aa,LaMassa:2013ab,LaMassa:2016aa} with an exposure time of $\sim$4.1\,ks. The Stripe 82X coverage is a combination of archival XMM data and data awarded to us from two observing campaigns, one covering $\sim 4.5$ deg$^2$ from AO10 \citep{LaMassa:2013aa} and the other covering 15.6 deg$^2$ from AO13 \citep{LaMassa:2016aa}.  CQ 4479 was specifically observed in the \citet{LaMassa:2013aa} dataset with a pixel scale of 4.1\arcsec and individual European Photon Imaging Camera exposure field of view of 30\arcmin.

The spectral analysis (Peca et al. in preparation) using this data reveals 82 net counts after background subtraction in the Full (0.3-10 keV) band. The hardness ratio, defined as $HR=(H-S)/(H+S)$, where $H$ and $S$ are the net count rates in the Hard (2-10 keV) and Soft (0.3-2 keV) bands, can be used to identify obscuration when assuming a power law with an absorber model \citep{LaMassa:2016ab}. CQ4479 has a hardness ratio of $HR \sim -0.8$.  The negative HR, observed at $z$ = 0.405, is typical of unobscured AGN \citep{LaMassa:2016ab}.

\begin{figure}
	\begin{center}
		\includegraphics[width=0.5\textwidth]{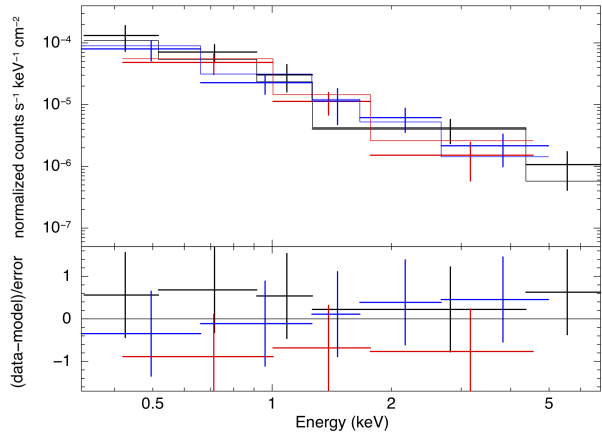}\hfil
		\caption{Top: deconvolved spectrum (crosses) and best-fit model (solid line) observed with XMM-Newton. PN, MOS1 and MOS2 cameras are plotted in black, red, and blue, respectively. Bottom: residuals are shown. The spectrum is rebinned for visual clarity.}
		\label{fig:specxmm}
	\end {center}
\end{figure}

The X-ray spectrum (Figure \ref{fig:specxmm}) is modeled with an absorbed power law plus a galactic absorption of $N_{\rm H}=5\times 10^{20}$ cm$^{-2}$. The best-fit model yields an unobscured ($N_H< 10^{21}$ cm$^{-2}$) AGN, with a steep power-law photon index, $\Gamma=2.4_{-0.3}^{+0.4}$ ($N_H$ and errors at 90\% confidence level), from which we can infer a higher accretion rate than commonly observed at this redshift 
\citep[e.g.,][]{Shemmer:2008aa,Risaliti:2009aa}. The derived absorption-corrected and k-corrected X-ray luminosity is $L_{\rm 2-10\,keV} = 3.4^{+0.15}_{-0.19} \times 10^{43}$ erg s $^{-1}$, with 90\% error confidence levels.  To explore how sensitive our fit luminosity is to fitting procedure, we also fit the X-ray spectrum with a fixed upper limit for $N_H$ and $\Gamma$ left as a free parameter.  This fit results in a steeper power law $\Gamma=2.7$ and a lower deabsorbed intrinsic AGN luminosity of $L_{\rm 2-10\,keV} = 2.7 \times 10^{43}$ erg s $^{-1}$.  Finally, we fix $\Gamma=2.7$, leaving $N_H$ as a free parameter and estimate a luminosity of $L_{\rm 2-10\,keV} = 3.9 \times 10^{43}$ erg s $^{-1}$, also recovering an unobscured quasar solution with an upper limit of $N_H< 10^{21}$ cm$^{-2}$. All fits are consistent with an unobscured AGN system at the heart of CQ 4479 (Table~\ref{tab:xrayfitting}); however, the individual fits are limited in their ability to converge due to our net source count of 82 photons.  Therefore we proceed in the analysis with the best-fit model yielding $L_{\rm 2-10\,keV} = 3.4^{+0.15}_{-0.19} \times 10^{43}$ erg s $^{-1}$.

\begin{deluxetable}{lcrr}
 \tablecaption{CQ 4479 X-Ray Fitting Parameters
 \label{tab:xrayfitting}}
 \tablehead{\colhead{Model} & \colhead{$\gamma$}& \colhead{$N_{\rm H}$ (cm$^{-2}$)} } 
 \startdata
 Free & $2.4_{-0.3}^{+0.4}$ & $1.1\pm0.1<10^{21}$   \\ 
  $\gamma$-fixed & [2.4] & $0.7\pm0.1<10^{21}$  \\ 
  $N_{\rm H}$-fixed & 2.7 & [$1.1\pm0.1<10^{21}$]  \\ 
 $\gamma$-fixed & [2.7] & $1.1\pm0.1<10^{21}$  
 \enddata
 \tablecomments{X-ray spectrum fit parameter results from an absorbed power-law fit corrected for galactic absorption.  When both parameters are left free, or one of each parameter is fixed using previous fit results, we consistently find an upper limit solution to $N_{\rm H}$ characteristic of an unobscured quasar.}
 \end{deluxetable}

\begin{figure*}
\begin{center}
\includegraphics[width=0.7\textwidth]{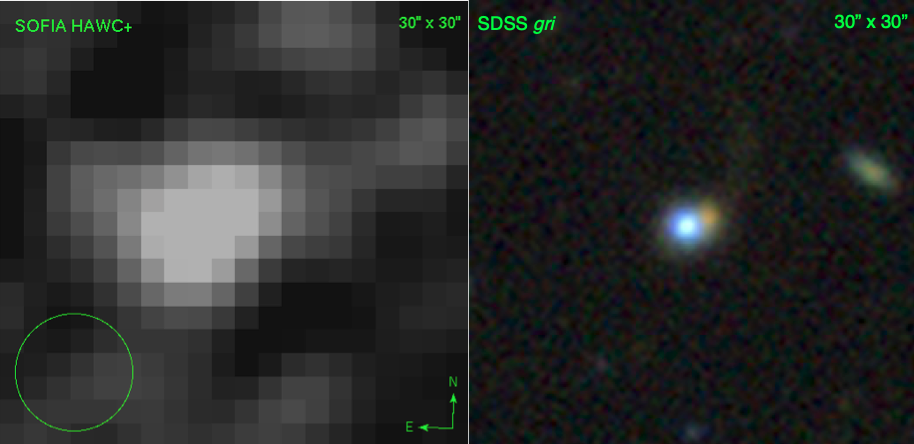}
\caption{Left:  SOFIA HAWC+ Band C observation of CQ 4479 centered at the target coordinates of R.A. J2000 of 1h 40m 40.71s and Decl. J2000 of 0$\degree$ 17$\arcmin$ 58.17$\arcsec$.  For reference, the SOFIA HAWC+ Band C (89 $\micron$) PSF FWHM is the circle in the lower left (7.8$\arcsec$).  At z = 0.405, the spatial scale is 5.413 kpc/\arcsec \citep{Wright:2006aa}. Each pixel is 1.55$\arcsec$, subtending a linear width of 8.4 kpc.  Right:  SDSS $gri$ composite image.  Note the far-IR image is resolved and asymmetric. The quasar point source is visible in the $gri$ composite, with resolved and asymmetric host galaxy light seen in the redder optical bands and the derived 89 micron flux provided alongside the archival photometry in Table \ref{tab:inputphoto}.}
\label{fig:image}
\end{center}
\end{figure*}


\subsection{Far-ultraviolet to Far-infrared Photometry}\label{sec:photometry}

Far-ultraviolet (FUV) to far-infrared (FIR) observations are obtained from the Stripe 82X catalog \citep{Ananna:2017aa}.  The target has FUV and NUV coverage from GALEX \citep{Martin:2005aa}, and optical coverage from the coadded SDSS catalogs \citep{Jiang:2014aa,Fliri:2015aa}.  In the infrared, Stripe 82X was observed with the Spitzer Space Telescope at 3.6 and 4.5~$\micron$ \citep{Papovich:2016aa,Timlin:2016aa} as well as the WISE 3.4, 4.6, 12, and 22~$\micron$ bands \citep{Wright:2010aa}.  Herschel/SPIRE observations using 250, 350, and 500~$\micron$ bands are also available for the Stripe82X field from the Herschel Stripe82 Survey \citep[HerS;][]{Viero:2014aa}.  For more information on the Stripe 82X catalog and its counterpart identification protocol, please see \citet{LaMassa:2016aa} and \citet{Ananna:2017aa}.

		
\subsection{SOFIA HAWC+ Observations}
Due to the large gap in spectral coverage from 22 to 250 $\micron$, crucial for measuring the dust-obscured star formation rate (SFR) and the AGN contribution to the mid-IR, new observations  using SOFIA/HAWC+ \citep{Dowell:2013aa,Smith:2014aa} were taken using band C at 89 \micron\;(PI: A. Kirkpatrick, PID: 07-0096).  CQ 4479 was observed for $\sim$7223 s on 2019 September 7 during SOFIA Cycle 7 using the HAWC+ total intensity OTFMAP mode with a Lissajous pattern in nominal water vapor conditions. At this redshift, each 1.55$\arcsec$ pixel subtends a linear width of 8.4 kpc.  The total field of view of HAWC+ in this setup is 4.2\arcmin\; by 2.7\arcmin, with an FWHM of 7.8$\arcsec$.  Following data acquisition, the observations were reduced using the HAWCDPR PIPELINE \citep{Harper:2018aa}. Flux calibration was performed using an average reference calibration factor across the detector. Source extraction was performed using the CRUSH V. 2.50-2 {\tt detect} routine \citep{Kovacs:2006aa,Kovacs:2008aa} in the faint object mode, yielding a flux measurement of $75.42\pm14.2$ mJy in a 7.8\arcsec\; FWHM beam (SNR = 5.31).  
The SOFIA HAWC+ band C image is shown in Figure \ref{fig:image}. 

\begin{figure}
\epsfig{file=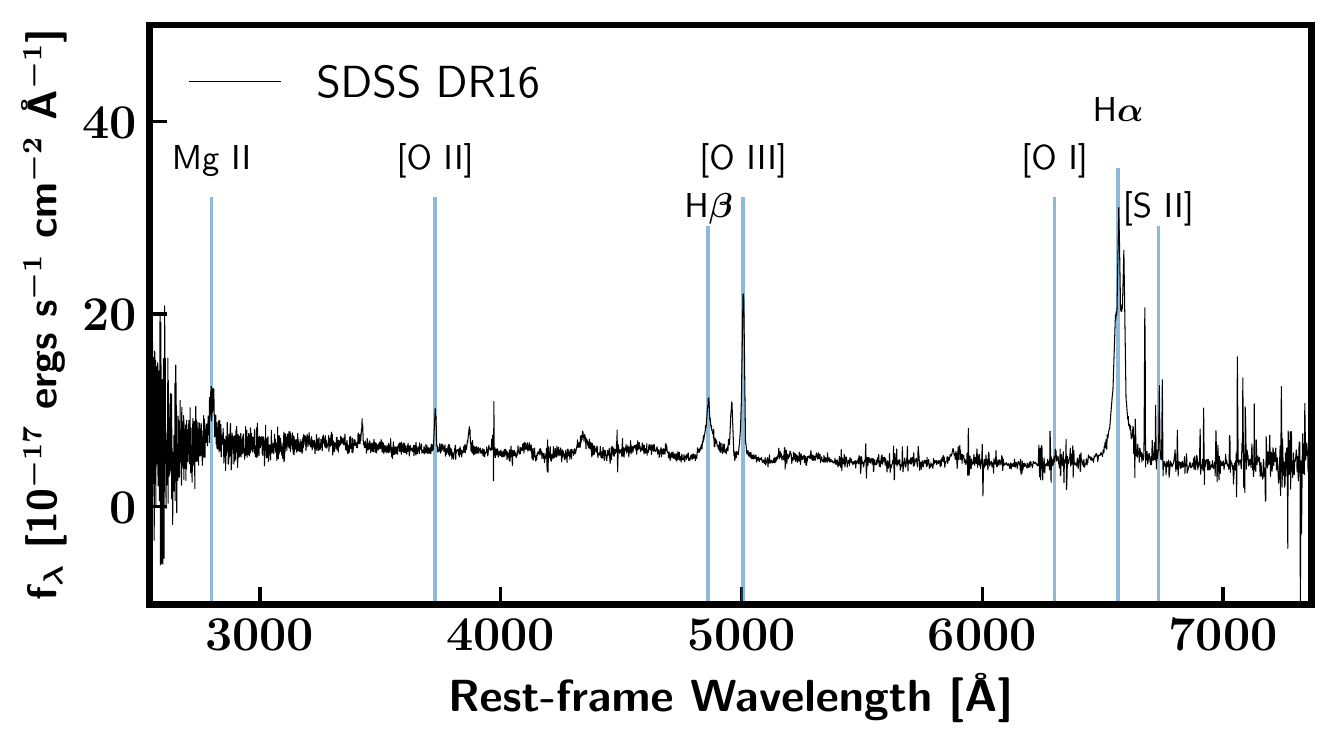,width=0.47\textwidth,angle=0}
\caption{The SDSS spectrum for CQ 4479 with emission lines annotated. Broad Mg~{\sc ii}, H$\beta$, and H$\alpha$ lines are clearly visible, consistent with the classification of a Type-1 quasar.}
\label{fig:opspec}
\end{figure}


\section{Results}\label{sec:methods}


\subsection{Optical Emission Lines}\label{sec:optlines}
\begin{figure}[ht!]
    \centering
    \includegraphics[width=0.49\textwidth]{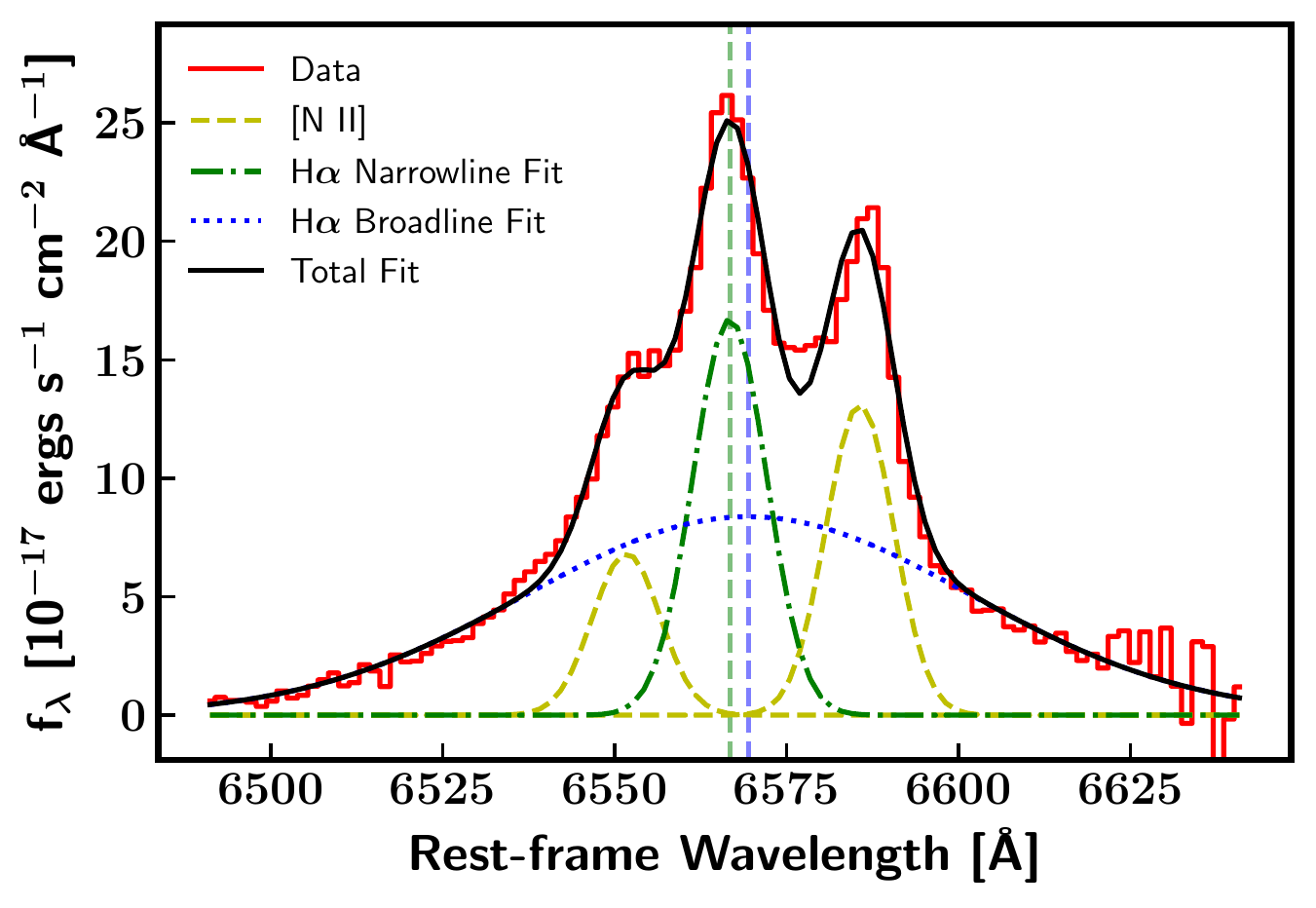}
    \includegraphics[width=0.49\textwidth]{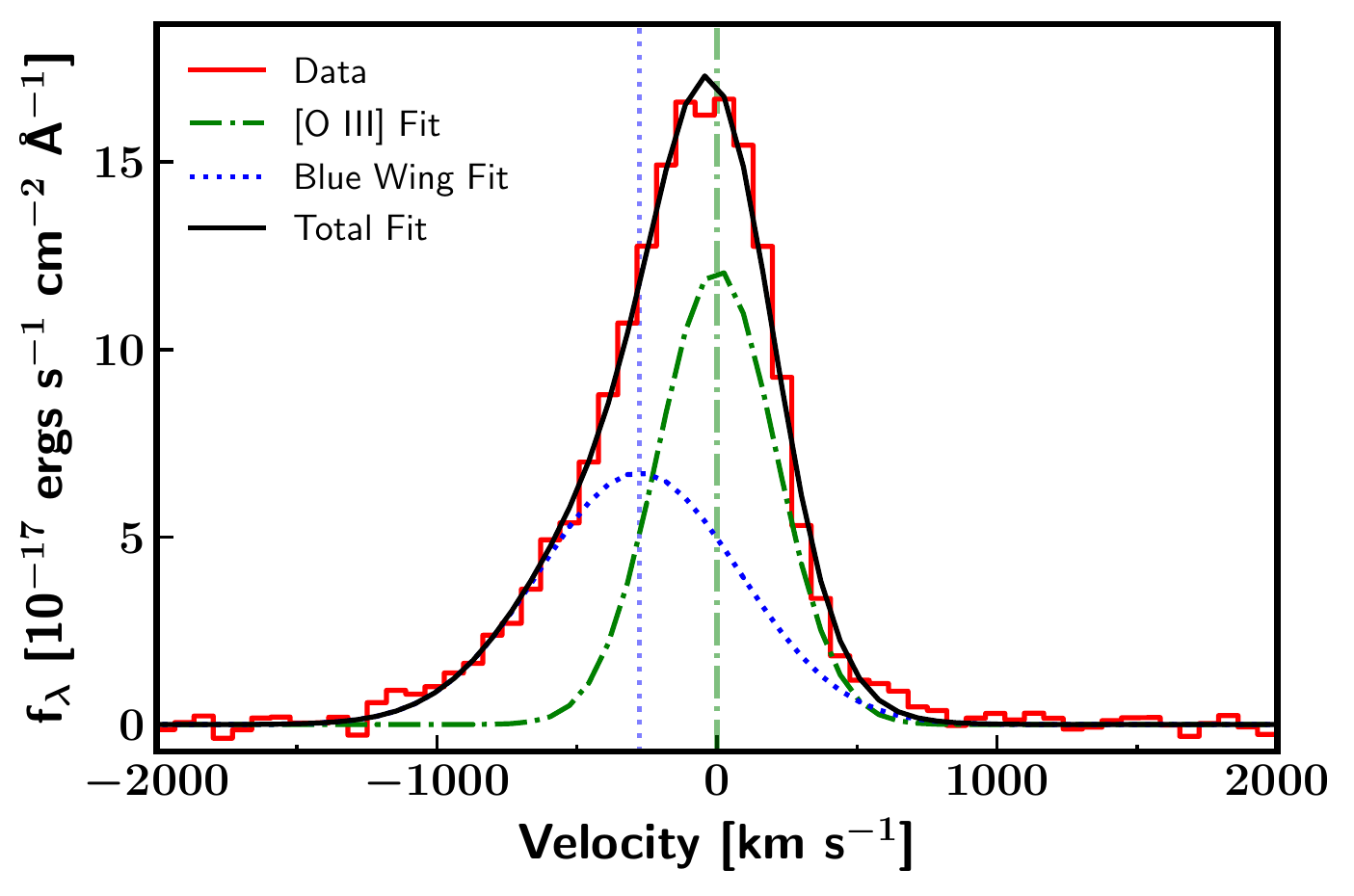}
    \caption{Top: total fit (black) to the H$\alpha$-[N II] emission line region is plotted in comparison to the SDSS DR16 spectrum (red stepped), including the H$\alpha$ broadline component (blue dotted), the H$\alpha$ narrow-line component (green dotted--dashed), and the [{\sc N ii}] doublet (yellow dashed).  Bottom: Observed [{\sc O iii}] emission line flux (red stepped) in comparison to the narrow component fit (green dotted--dashed),  blue wing fit (blue dotted), and total fit (black solid).  [{\sc O iii}] shows the potential signature of a outflow.}
    \label{fig:HA_fit}
\end{figure}

We analyze the optical emission lines using the SDSS-IV spectrum from Data Release 16 \citep{Ahumada:2019aa}.  The SDSS spectrum (Figure \ref{fig:opspec}) exhibits classic Type-1 quasar emission, with broad Mg~{\sc ii}, H$\beta$, and H$\alpha$ emission lines.
We measure the black hole mass by fitting a multicomponent Gaussian to the H$\alpha$/[N~{\sc ii}] complex at $\sim$6563\,\AA\ (Figure \ref{fig:HA_fit}), which consists of a broad and narrow H$\alpha$ component as well as the [N~{\sc ii}] doublet. We allow the normalization, central wavelength, and line widths to be independent parameters. We attribute the broadening behavior of the H$\alpha$ line as being due to the gas moving under the gravitational influence of the black hole. We use the broad H$\alpha$ component to determine the black hole mass with the relation, 
\begin{equation}
    \left(\frac{M_\bullet}{M_\odot}\right) =1.3\times10^6\left(\frac{L_{\rm H\alpha}}{10^{42}{\rm erg^{-1}}}\right)^{0.57}\left(\frac{\rm FWHM_{H\alpha}}{1000\,{\rm km^{-1}}}\right)^{2.06},
\end{equation}
where L$_{H\alpha}$ and FWHM$_{H\alpha}$ are the luminosity and FWHM of the broad H$\alpha$ component \citet{Greene:2005aa}. This yields $M_\bullet = 2.41\pm0.26\times10^7\,$M$_\odot$.

We also fit the H$\beta$ and [O {\sc iii}] emissions with a model consisting of broad and narrow Gaussian components. The fit to the [O {\sc iii}] emission line is shown in the bottom panel of Figure \ref{fig:HA_fit}. The narrow [O {\sc iii}] component has an FWHM of 492\,km s$^{-1}$, consistent with other Type 1 quasars \citep{Schmidt:2018aa}. The [O {\sc iii}] emission is asymmetric, with a blue wing centered at $\Delta v = -202.41$\,km s$^{-1}$ from the [O {\sc iii}] core and an FWHM of 845\,km s$^{-1}$, well within the typical range reported in the literature \citep{Bian:2005aa,Boroson:2005aa,Schmidt:2018aa}.  We compare the blue wing properties to the other optical emission lines in Table~\ref{tab:sdsslinefits} and find no other component with a similar offset or FWHM estimate. 

\begin{figure}
\begin{center}
\includegraphics[width=0.49\textwidth]{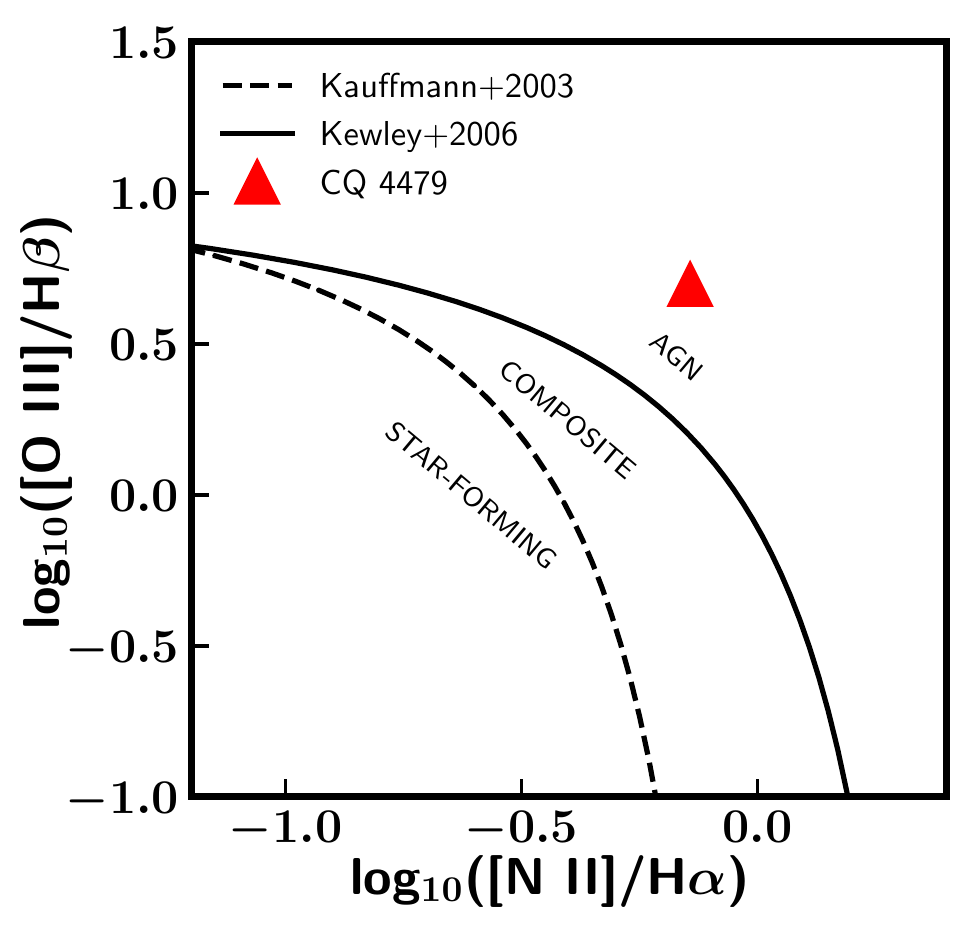}
\caption{The location of CQ 4479 (red triangle) in relation to the classical BPT diagram classification regions.  Optical emission line values are derived from the SDSS DR16 spectrum shown in Fig. \ref{fig:opspec}, with errorbars within the size of the symbol. Here we only consider the [O III] emission from the narrow-line component, as we want to characterize the host galaxy without potential outflow contamination.}
\label{fig:bpt}
\end{center}
\end{figure}

\begin{deluxetable}{lcrrrr}
 \tablecaption{CQ 4479 Photometry
 \label{tab:inputphoto}}
 \tablehead{\colhead{Filter} & \colhead{$\lambda_{obs}$ ($\micron$)}& \colhead{Flux}& \colhead{Flux Err}& \colhead{Flux Units} } 
 \startdata
 FUV & 0.152 & 17.5 & 1.1 & $\mu$Jy &\\ 
 NUV & 0.227 & 31.8 & 1.1 &$\mu$Jy &\\ 
 $u$ & 0.354 & 59.2 & 0.3 &$\mu$Jy &\\ 
 $g$ & 0.477 & 89.5 & 0.1 &$\mu$Jy &\\ 
 $r$ & 0.623 & 122.0 & 0.2 &$\mu$Jy &\\ 
 $i$ & 0.762 & 154.0 & 0.4 &$\mu$Jy &\\ 
 $z$ & 0.913 & 203.1 & 1.3 &$\mu$Jy &\\
 jVHS & 1.25 & 234.5 & 5.1 &$\mu$Jy &\\
 hVHS & 1.64 & 338.8 & 7.1 &$\mu$Jy &\\
 kVHS & 2.15 & 527.6 & 13.8 &$\mu$Jy &\\
 WISE1 & 3.40 & 715.8 & 17.8 &$\mu$Jy &\\
 IRAC1 & 3.56 & 807.3 & 5.9 &$\mu$Jy &\\
 IRAC2 & 4.51 & 1.03 & 0.01 &mJy &\\
 WISE2 & 4.60 & 1.00 & 0.02 &mJy &\\
 WISE3 & 12. & 3.11 & 0.2 &mJy &\\
 WISE4 & 22. & 9.38 & 0.1 &mJy &\\
 HAWC+ C & 89. & 75.4 & 14.2 &mJy &\\
 SPIRE1 & 250. & 51.5 & 10.1 &mJy &\\
 SPIRE2 & 350. & 20.8 & 10.2 &mJy &\\
 SPIRE3 & 500. & 7.2 & 10.8 & mJy
 \enddata
 \tablecomments{Observed and archival photometry with associated errors for each wave band used in the broadband SED fitting described in Section \ref{sec:fitting}.}
 \end{deluxetable}

Finally, we use the narrow components of the optical emission lines to confirm the source of ionizing radiation, and CQ 4479 has clear signatures of a Type-1 AGN (Figure \ref{fig:bpt}). The ratio of [N~{\sc ii}]/H$\alpha$ vs. [O {\sc iii}]/H$\beta$ falls securely in the AGN region of the BPT diagram \citep{Baldwin:1981aa},  strongly indicating a high ionization state due to radiation from an AGN \citep{Kauffmann:2003aa,Kewley:2006aa}.


\subsection{Broadband SED Fitting}\label{sec:fitting}
We use two SED fitting packages to decompose the stellar, dust, and AGN contributors to the UV-FIR photometry (Table \ref{tab:inputphoto}).  We choose {\tt SED3FIT} \citep{Berta:2013aa} and {\tt CIGALE} \citep{Burgarella:2005aa,Boquien:2019aa}. Both codes employ an energy balance requirement to ensure that the dust luminosity is consistent with the energy provided by obscured stellar and AGN radiation. We chose to employ two codes to provide a consistency check. {\tt SED3FIT} provides a direct estimation of the dust temperature, while {\tt CIGALE} provides the faster run time and larger range of model libraries, enabling better systematic error estimation.

\begin{deluxetable*}{lcrrrrr}
 \tablecaption{CQ 4479 Optical Emission Line Fit Parameters
 \label{tab:sdsslinefits}}
 \tablehead{\colhead{Line} & \colhead{$\lambda_{fit}$}& \colhead{Flux}& \colhead{FWHM}& \colhead{Luminosity}& \colhead{vel$_{relative}$} \\
            \colhead{} & \colhead{$\AA$}& \colhead{(10$^{-17 }$erg s$^{-1}$ cm$^{-2}$)}& \colhead{(km s$^{-1}$)}& \colhead{log$_{10}$(erg s$^{-1}$)}& \colhead{(km s$^{-1}$)} 
 } 
 \startdata
 H$\beta_{Broad}$ & 4863 & 101.96$\pm$0.90 & 2287 &  41.77$\pm$0.01 & +151.7 \\
 H$\beta_{Narrow}$ & 4862 & 18.50$\pm$0.61 & 559 &  41.03$\pm$0.01 & +77.6 \\
 $[$O III$]$ & 5008 & 92.09$\pm$0.27 & 492 & 41.73$\pm$0.01 &  +75.5\\
 $[$O III$_{Blue Wing}]$ & 5003 & 87.51$\pm$0.40 & 845 & 41.70$\pm$0.01 & -202.41\\
  $[$N II$_a]$ & 6551 & 55.93$\pm$1.05 & 531 & 41.51$\pm$0.01 & +85.3\\
 H$\alpha_{Broad}$ & 6569 & 439.54$\pm$2.41 & 3475 & 42.40$\pm$0.01 & +219.9 \\
  H$\alpha_{Narrow}$ & 6566 & 148.03$\pm$0.50 & 574 & 41.93$\pm$0.01 & +97.9 \\
  $[$N II$_b]$ & 6585 & 106.56$\pm$0.45 & 526 & 41.79$\pm$0.01 & +18.0
 \enddata
 \tablecomments{Best fit parameters derived from the SDSS optical emission line fits described in Section \ref{sec:optlines}.  Relative velocity is calculated with respect to the target's spectroscopic redshift, with positive values denoting redshift and negative values denoting blueshift relative to the host.  The blue wing component of the [O III] emission line occupies a distinct velocity space with respect to other lines.}
 \end{deluxetable*}

{\renewcommand{\arraystretch}{1.5}
\begin{deluxetable*}{l|c|c|c|c}
\tablecolumns{3} 
 \tablecaption{Parameter Fits to CQ 4479 Using {\tt SED3FIT} and {\tt CIGALE}
 \label{tab:fitresults}}
 \tablehead{\colhead{Input Parameters} & \colhead{{\tt SED3FIT w/ SOFIA}} &\colhead{{\tt SED3FIT w/o SOFIA}} & \colhead{{\tt CIGALE w/ SOFIA}} & \colhead{{\tt CIGALE w/o SOFIA}} } 
 \startdata 
Stellar Population & \citet{Bruzual:2003aa} &\citet{Bruzual:2003aa} & \citet{Bruzual:2003aa} & \citet{Bruzual:2003aa} \\ 
Dust & \citet{da-Cunha:2008aa} &\citet{da-Cunha:2008aa} & \citet{Draine:2014aa}\tablenotemark{a} & \citet{Draine:2014aa}\tablenotemark{a}\\
AGN Model & \citet{Feltre:2012aa} &\citet{Feltre:2012aa} & \citet{Fritz:2006aa} & \citet{Fritz:2006aa} \\
Initial Mass Function & \citet{Chabrier:2003aa} &\citet{Chabrier:2003aa} & \citet{Chabrier:2003aa} & \citet{Chabrier:2003aa} \\ 
\hline \multicolumn{1}{c|}{Output Parameters} &  &  &  &  \\ 
\hline SFR (M$_{\odot}$ yr$^{-1}$) & 78 $\pm$ 20 & 38.19 $\pm$ 2.26 & 95.55 $\pm$ 4.59 & 51.85 $\pm$ 2.68 \\ 
 M$_{*}$ (10$^{10}$ M$_{\odot}$) & 3.11 $\pm$ 0.2 & 3.01 $\pm$ 0.13 & 2.19 $\pm$ 0.37 & 2.11 $\pm$ 0.39 \\ 
 log$_{10}$(sSFR (yr$^{-1}$)) & -8.625 $\pm$ 0.05 & -8.89 $\pm$ 0.05 & -8.36 $\pm$ 0.1 & -8.60 $\pm$ 0.08\\
 $A_{v}$ & 0.61 $\pm$ 0.22 & ... & 0.90 $\pm$ 0.02 & 0.95 $\pm$ 0.06 \\
 $T_{\rm Dust}$ (K) & 59.15 $\pm$ 26.3 & 39.163 $\pm$ 14.73  &  ... & ... \\
 $L_{\rm IR} (8-1000\micron)$ (10$^{11}$ L$_{\odot}$)  & 6.76 $\pm$ 2.29 & 3.43 $\pm$ 1.54 & 8.28 $\pm$ 0.41 & 4.10 $\pm$ 0.28\\
 $M_{\rm Dust}	      $ (10$^{8}$ M$_{\odot}$)  & 2.00 $\pm$ 0.29 & 2.73 $\pm$ 0.86 & 5.04 $\pm$ 0.52 & 6.68 $\pm$ 0.27\\
 $M_{\rm Gas}\tablenotemark{b}$ (10$^{10}$ M$_{\odot}$)  & 2.00 $\pm$ 0.29 & 2.73 $\pm$ 0.86 & 5.04 $\pm$ 0.52 & 6.68 $\pm$ 0.27\\
 $\tau_{\rm Dep}$ (Gyr)         & 0.256                     & 0.71 & 0.526                   & 1.28 \\
 $f_{\rm AGN} (8-1000\micron)$ & 9.7\% & 28\% & 10\% & 20\%\\ 
 $\chi^2_r$ & 5.26 & 5.90 & 0.99 & 1.22
 \enddata
 \tablenotetext{a}{The \citet{Draine:2014aa} models are a modified version of the \citet{Draine:2007aa} available in the public {\tt CIGALE} distribution.}  
 \tablenotetext{b}{Gas masses are estimated assuming a dust-to-gas ratio of 1:100 derived from local observations \citep{Bohlin:1978aa}.}
\end{deluxetable*}
}

{\tt SED3FIT} is a publicly available SED fitting package built upon the public {\tt MAGPHYS} code \citep{da-Cunha:2008aa}, adding the capability to fit an AGN with the optical and IR components.  For each target, both packages construct model libraries using stellar models from \citet{Bruzual:2003aa} and IR dust models from \citet{da-Cunha:2008aa}.  In {\tt SED3FIT}, AGN models from \citet{Feltre:2012aa} are included across both optical and infrared fitting steps, and include both high and low viewing angle templates \citep[Type-1 and Type-2 AGN,][]{Urry:1995aa}.

\begin{figure*}
\begin{center}
\epsfig{file=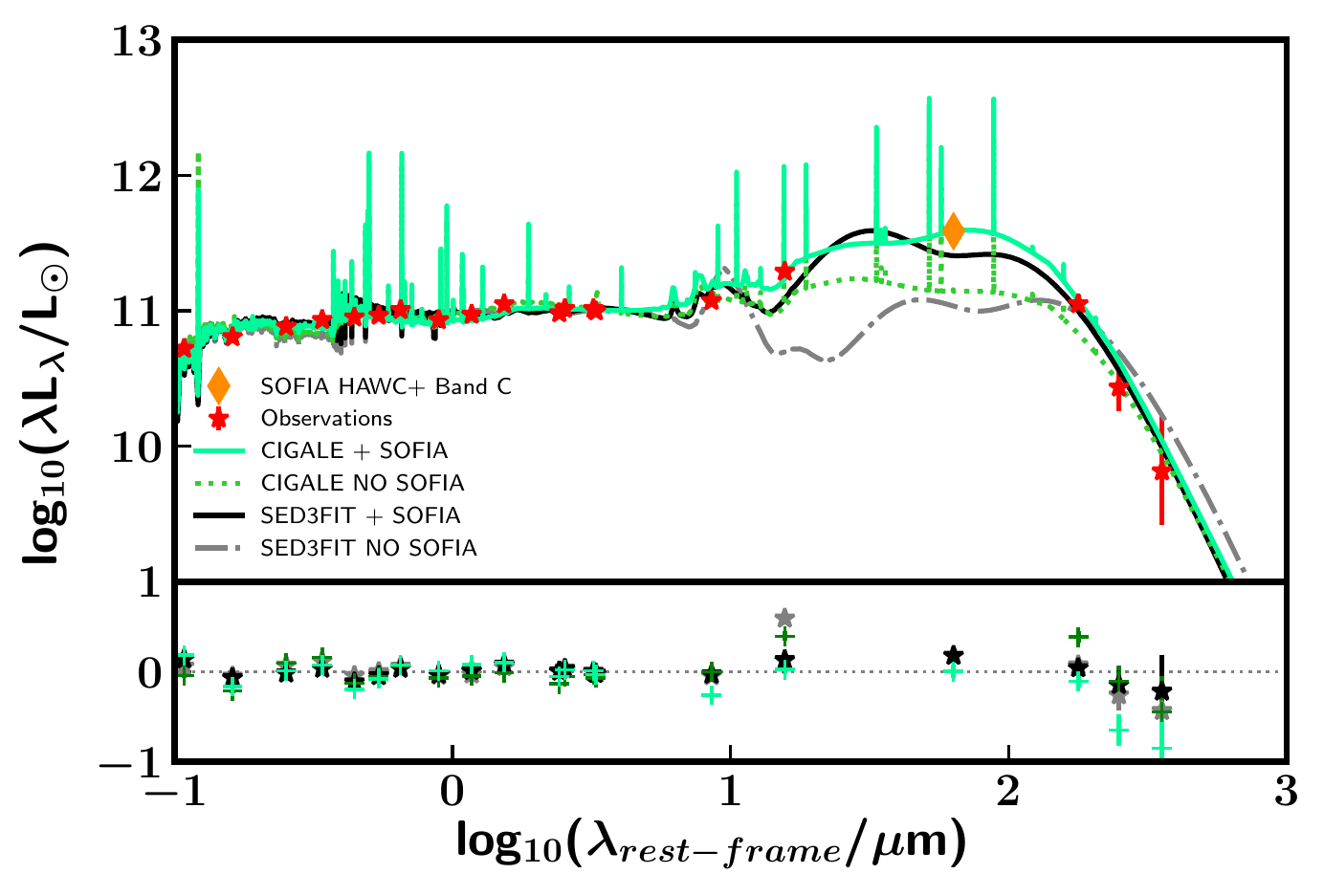,width=0.95\textwidth,angle=0}
\caption{Top: the CIGALE (green) and SED3FIT (black) estimated SEDs of CQ 4479 both with (solid) and without (dashed) the SOFIA 89 $\micron$ detection in orange.  The non-SOFIA photometric data are plotted as red star symbols. Bottom:  the fractional residual $\frac{f_{obs}-f{model}}{f_{obs}}$ with errors as the vertical lines of each fitting package for each band.  Color coding is identical in both panels.}
\label{fig:sedplot}
\end{center}
\end{figure*}

Code Investigating GALaxy Emission ({\tt CIGALE}) is an SED fitting package that fits galaxy SEDs from the FUV to radio wave bands. 
We use \citet{Bruzual:2003aa} stellar population models, \citet{Fritz:2006aa} AGN models, and both the original \citet{Draine:2007aa} and updated \citet{Draine:2014aa} dust models. Both dust models produce consistent results, so we only list the results using \citet{Draine:2014aa} in Table \ref{tab:fitresults}. {\tt CIGALE}'s built-in nebular emission lines are also included. We compare the best-fit {\tt CIGALE} and {\tt SED3FIT} SEDs in Figure \ref{fig:sedplot}.  

The fitted parameters from both SED packages, using and excluding the new SOFIA data, are cataloged in Table \ref{tab:fitresults}. The inclusion of the new SOFIA HAWC+ 89 $\micron$ photometry significantly changes the modeled FIR output of CQ 4479 and is an example of the powerful characterization possible with FIR detectors such as those on SOFIA or future telescope concepts such as Origins Space Telescope \citep{Meixner:2019aa}. When the SOFIA 89\,$\micron$ data is included in the fit, the IR luminosity (LIR) and SFR increase by a factor of two, while the AGN contribution to the LIR decreases by roughly the same amount. This drop in the AGN fraction is primarily due to the increase in the total infrared luminosity of the galaxy, rather than a change in the estimated amount of torus emission. That is, the fraction of AGN emission at 2.5-5.0 um changes very little in the fits with and without the SOFIA data (88\% and 83\% AGN fraction at these wavelengths, respectively).  Therefore the critical difference in the SED fit when including SOFIA data is from the dust component of the IR fit rather than the AGN.


The fits from both fitting methods match reasonably well, with $1 < \chi^2_r < 5$ for best fits. Both packages agree that the AGN contribution to the IR emission from 8 to 1000 $\micron$ is low at 10\%, and that the IR luminosity is strongly driven by SFRs of $\sim$90 M$_{\odot}$ yr$^{-1}$.  We also run an alternate series of fits using upper limits for the SPIRE points and recover SFR estimates within the errors.  We find a gas mass fraction ($M_{\rm gas}/M_{\rm stars+gas}$) of $50-70$\%, where the gas mass derived from the dust mass, assuming a gas-to-dust ratio of 100 \citep{Bohlin:1978aa}. In the analysis that follows, we use the SFR, $M_{\rm gas}$, and $M_\ast$ from the {\tt CIGALE} fit, given that the $\chi^2_r =$ 0.99, better than that from the {\tt SED3FIT} model.  

\section{Discussion}\label{sec:results}

Cold quasars are a population of Type-1 quasars with extreme SFRs and high gas masses \citep{Kirkpatrick:2020aa}. In the context of an evolutionary scenario when an unobscured quasar phase follows an obscured phase \citep[e.g.,][]{Hopkins:2006aa}, cold quasars would be caught in the act of transitioning, i.e. the AGN is dispersing the surrounding ISM in a process known as ``blowout." In the case of CQ 4479, a blue wing component is observed as a component of the [O III] emission, and likely arises from an outflow close to the AGN powering a blowout phase, although without spatially resolved spectroscopy, we cannot determine the extent of the narrow-line region. Unobscured quasars in general have lower SFRs than cold quasars \citep{Stanley:2015aa,Stanley:2017aa,Kirkpatrick:2020aa}, and if the unobscured quasar phase directly precedes the ``red and dead'' elliptical phase, CQ 4479 may be in the process of quenching due to AGN and high SF activity.


\subsection{Gas Depletion Timescale}
To characterize the state of quenching, we estimate the 
gas depletion time ($\tau_{\rm dep} = M_{\rm gas}$/SFR) to determine the likely future of CQ 4479.  We estimate that the gas depletion times are extremely short, $\lesssim$ 0.5 Gyr
, a factor of 2 or 6 shorter than local star-forming and quiescent galaxies \citep{Saintonge:2011ab}. This is the gas depletion timescale due to star formation alone, but it is possible for the AGN to drive molecular outflows that can also deplete the gas reservoir further \citep{Chen:2019aa,Herrera:2019aa,Herrera:2020aa}.
We therefore take 0.5\,Gyr as an upper limit for $\tau_{\rm dep}$.

\begin{figure}
\begin{center}
\includegraphics[width=0.49\textwidth]{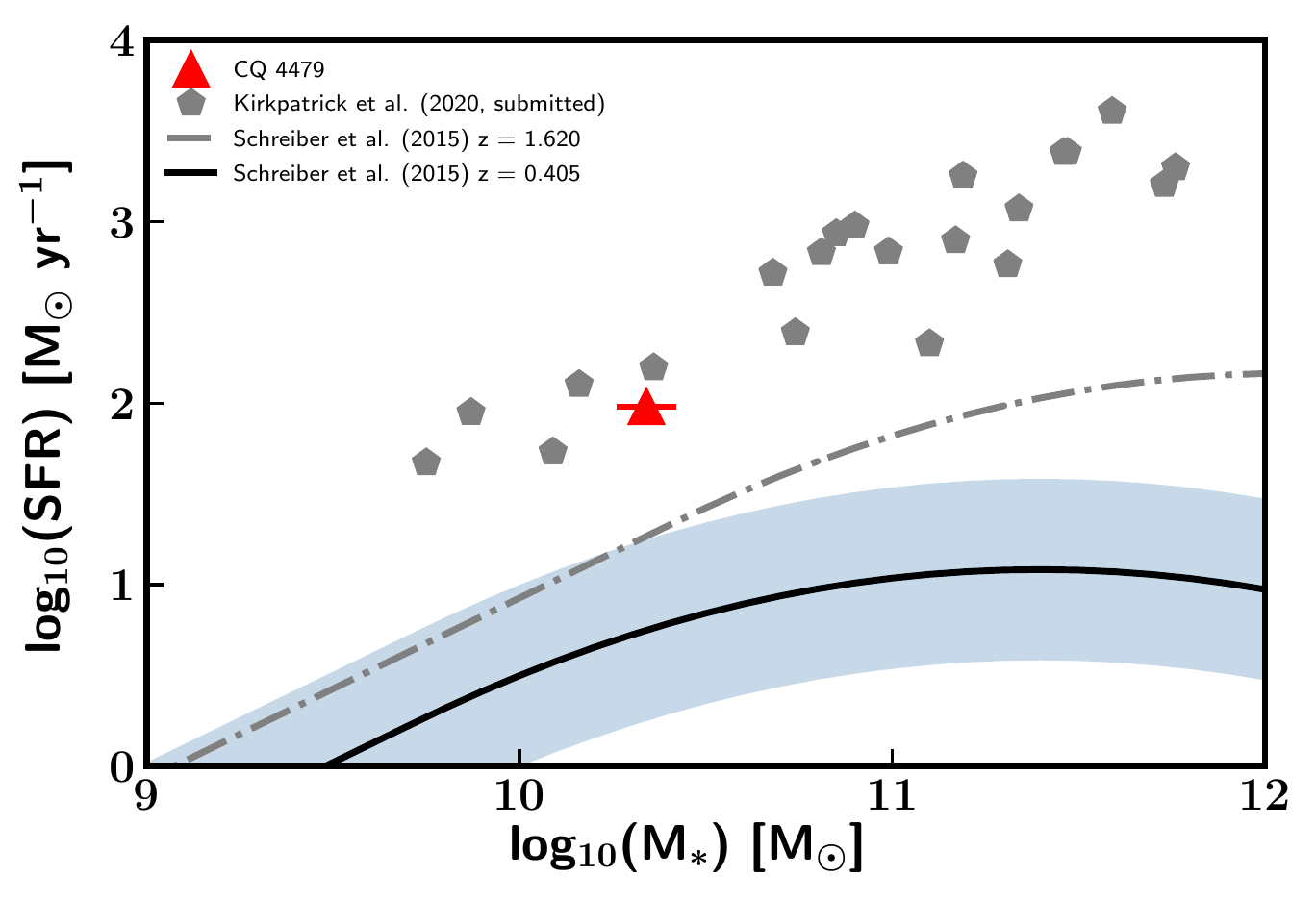}
\caption{The location of CQ 4479 (red triangle) and the parent sample (gray pentagons) in relation to the star-forming main sequence (SFMS) of galaxies at the target redshift (z = 0.405; black) and the SFMS at the median redshift of the total sample (z = 1.62; gray).  The blue shaded region covers the range of SFRs within a factor of 3 of the field-aggregate SFMS relation of \citet{Schreiber:2015aa} at $z$ = 0.405.  Galaxies above the blue region are classified as starburst galaxies. CQ 4479 sits at the low M$_{*}$ end among the cold quasar sample.}
\label{fig:sfms}
\end{center}
\end{figure}

The short gas depletion timescale is consistent with its location on the SFR--stellar mass diagram (Figure \ref{fig:sfms}), which would classify this galaxy as a star-bursting galaxy, nearly a factor of 30 above the star-forming main sequence (SFMS) prediction for SFR activity for a galaxy of this stellar mass \citep{Schreiber:2015aa}. 

We compare CQ 4479 with the primary cold quasar sample from \citet{Kirkpatrick:2020aa} at $z\sim0.5-3$. The full cold quasar sample also lies in the starbursting regime according to the SFR and redshift \citep{Schreiber:2015aa}. Quenching has not yet begun in CQ 4479, based on its location in the starburst regime of the SFMS diagram, although the short $\tau_{\rm dep}$ reveals it may be imminent.

\begin{figure}
\begin{center}
\includegraphics[width=0.49\textwidth]{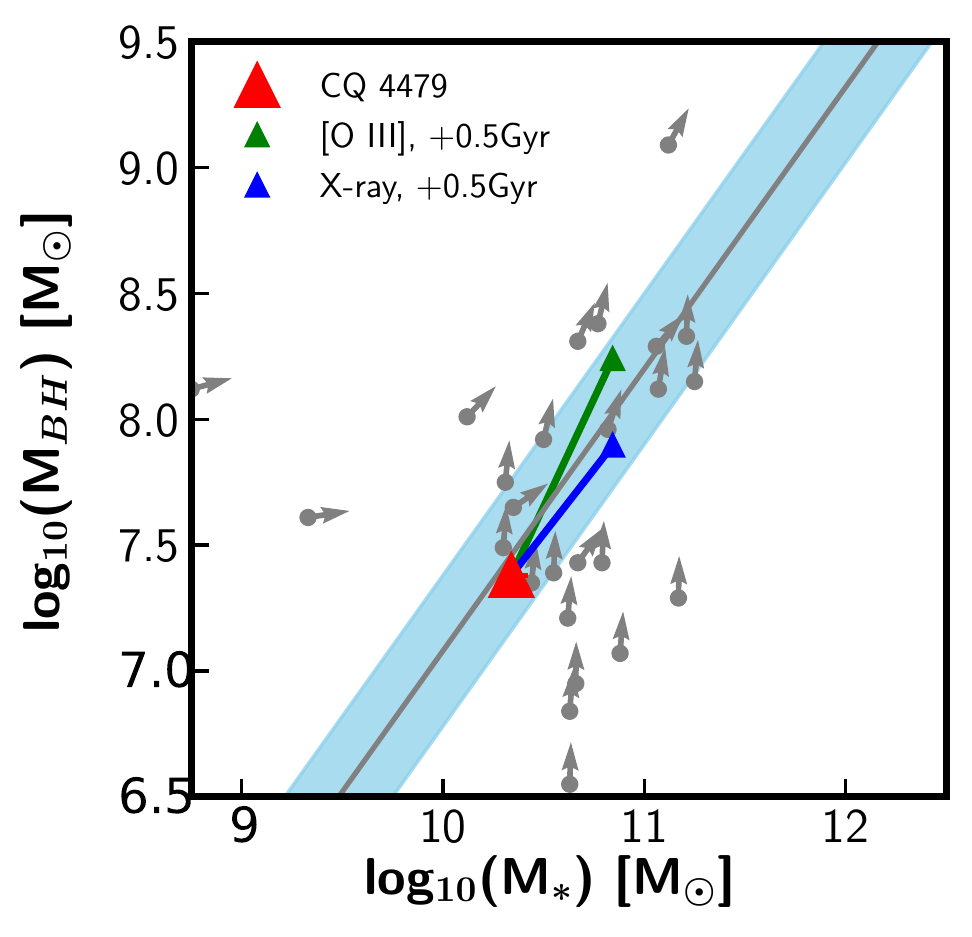}
\caption{CQ 4479 (red triangle) in relation to the empirical black hole--stellar mass relation \citep{Haring:2004aa}. For comparison, we include the low z ($z < 1$) sample of \citet{Sun:2015aa}.  Arrows indicate the path each galaxy is taking on this plot as a function of specific BH accretion rate ($\dot M_\bullet/M_\bullet$) and sSFR (SFR/$M_\ast$).  Green and blue points indicate the estimated final state of CQ 4479 after 0.5 Gyr (the upper limit on $\tau_{\rm dep}$) when using an [O {\sc III}] or X-ray defined SMBH accretion rate, respectively.}
\label{fig:flowdiagram}
\end{center}
\end{figure}



\subsection{Fading X-Ray Emission?}
Although the IR emission is dominated by star formation, the [O {\sc iii}] luminosity, $L_{\rm[O\;III]}$, can be used to provide an independent estimate of the bolometric luminosity due to the AGN using $L_{\rm bol}=3500\times L_{\rm [O\;III]}$, appropriate for [{\sc O iii}] emission uncorrected for extinction \citep{Heckman:2004aa}. We measure the [O III] line flux of both the core and the blue wing and obtain $L_{\rm [O III]} = (1.06 \pm 0.02) \times10^{42}$\,erg s$^{-1}$.  To derive a bolometric luminosity, we only consider the narrow-line component $L_{\rm [O\;III]} = (5.46 \pm 0.02) \times10^{41}$\,erg s$^{-1}$, yielding  $L_{\rm bol}^{\rm [O\;III]} = (1.91^{+2.67}_{-1.11}) \times 10^{45}$\,erg s$^{-1}$. Due to the high estimated SFR estimated from the SED fit of CQ 4479, part of the [{\sc O iii}] emission may arise from star formation. However, based on its location in the BPT diagram (Figure \ref{fig:bpt}), \citet{kauffmann:2009aa} predict that at most 10\% of the $L_{\rm [O III]}$ is due to star formation.  To verify this hypothesis, we independently estimate the component of $L_{\rm [O\;III]}$ expected from a star-forming galaxy hosting 95.5 M$_{\odot}$ yr$^-1$ of SF using the L$_{O\;III-H\beta}$--SFR relation of \citet{Drozdovsky:2005aa}, correcting to a Chabrier IMF, and assuming our measured [O III]/H$\beta$ ratio from the SDSS spectrum.  We estimate that SF may be responsible for between 100\% and 100\% of the [O III], a poor constraint due to the several order of magnitude dispersion in the [O III]/H$\beta$--SFR correlation at this redshift \citep{Drozdovsky:2005aa}. However, our target remains consistent with the lower bound of this correlation.

CQ 4479 has a hard X-ray luminosity of $L_{2-10\,\rm keV} =3.4\times10^{43}$\,erg s$^{-1}$. From $L_{2-10\,\rm keV}$, we derive $L_{\rm bol}^{X}=
(6.86\pm3.95)\times10^{44}$\,erg s$^{-1}$, using the relation derived in \citet{Lusso:2012aa} from a sample of $2-10$\,keV detected AGN with spectroscopic redshifts. This $L_{\rm bol}^{X}$ is lower than [{\sc O iii}], though they may be consistent at the extreme ends of range allowed by the uncertainties. 
Thus, CQ 4479 is underluminous in X-ray emission compared with the optical prediction.

As X-ray emission is not included in either of our SED fits, we use the 6\,$\mu$m emission of the SED fit's AGN component to calculate the expected X-ray emission, following the relation in \citet{Stern:2015aa}. We assume the torus model from the best-fit {\tt CIGALE} model to calculate $\nu L_{6\,\mu{\rm m}} = 1.91\times10^{44}\,$erg s$^{-1}$, which gives an estimated $L_{\rm 2-10\,keV}=6.76\times10^{43}$\,erg s$^{-1}$, twice as high as observed.  This behavior has been previously observed in red quasars \citep{Urrutia:2012aa,Glikman:2017aa}, a subpopulation of AGN with near Eddington-limit accretion rates yet whose X-ray spectra are best fit by an absorbed power law.  Additionally, this has been observed in hot dust-obscured galaxies \citep[hot DOGs;][]{Ricci:2017aa}, which are X-ray deficient subpopulations of quasars with FIR spectra that peak at much shorter wavelengths, $\lambda$ $\sim$ 20 $\micron$. While the FIR peaks we find are at cooler temperatures, to clarify the exact relationship between CQ4479 and hot DOGS would require further FIR observations.

The X-ray luminosity for CQ 4479 predicted by $\nu L_{6\,\mu{\rm m}}$ is more consistent with the higher levels of emission determined by optical properties. The best-fit {\tt CIGALE} and {\tt SED3FIT} AGN models can likewise be integrated to determine the bolometric luminosity of the AGN. Interestingly, these two fits give discrepant results. The {\tt SED3FIT} AGN component has a bolometric luminosity of $(1.65\pm0.02)\times 10^{45}$\,erg s$^{-1}$, while {\tt CIGALE} estimates $L_{\rm bol}=(6.78\pm0.03)\times10^{44}$\,erg s$^{-1}$. {\tt SED3FIT} is consistent with the [{\sc O iii}]-derived $L_{\rm bol}$ while {\tt CIGALE} is consistent with the X-ray-based $L_{\rm bol}$. This is most likely due to the differing amounts of AGN emission each model predicts in the optical, and illustrates the sensitivity of AGN derived properties to the particular SED fitting code employed. $L_{\rm bol}$ derived from SED packages has not been as well-calibrated in large samples as using $L_X$, $L_{\rm [O\;III]}$, and $\nu L_6$, so we only discuss the latter below.

The weaker X-ray emission than predicted by $L_{\rm [O\;III]}$ and $\nu L_6$ hints that accretion onto CQ 4479 is declining at the time of observation \citep{Sartori:2018aa,Ichikawa:2019aa}, or there X-ray emission is obscured and the spectrum cannot constrain $N_H$ accurately. Physically, the X-ray emission arises from the corona, which has the closest proximity to the accretion disk. It is therefore the most instantaneous measure of the accretion rate. Next in proximity is the infrared emission arising from the torus, typically on scales of $0.1-10$\,pc. Finally, the [O {\sc iii}] emission arises from the narrow-line region, typically $\sim100-1000$\,pc from the central BH. The size of the narrow-line region varies greatly from galaxy to galaxy, and has been found to span up to 20 kpc \citep[e.g.,][]{Hainline:2016aa}. It is the most extended emission in AGN, and therefore traces the accretion rate of the AGN on the longest timescales. CQ 4479 is a particularly tidy case, where $L_{\rm bol}$ increases as the physical tracer increases in distance from the AGN, pointing to a scenario when the accretion was higher in the past and is steadily decreasing \citep{Harrison:2017aa}. This AGN may be caught in the act of turning off, supporting the scenario where cold quasars are in a special, short-lived transition phase.  It is important to note the uncertainties in the bolometric correction, especially in the X-ray, and the X-ray luminosity estimated might not be the unabsorbed luminosity.

Alternatively, it is important to consider that the broad emission lines are driven by far-UV ionizing radiation that, like the corona, also comes from the inner disk. With BLR reverberation time delays of weeks it is difficult to explain how CQ 4479 would be underluminous in X-rays but still have a luminous BLR.  Previous work has generally suggested they are rapidly accreting \citep[e.g.][]{Brandt:2000aa} and wind-dominated systems \citep[e.g.][]{Gibson:2008aa}. Broad absorption line quasars (BALQSOs) are generally X-ray weak and fit this picture \citep{Boroson:2002aa}; however, the link between Eddington ratio and X-ray weakness remains under investigation \citep[e.g.][]{Vito:2018aa}.


\subsection{Predicting the Future of CQ 4479}
We can predict CQ 4479's future stellar mass and black hole mass growth. First, we calculate the black hole accretion rate, $\dot M_\bullet$, using the formula:
\begin{equation}
    \dot M_\bullet = \frac{(1-\eta)L_{\rm bol}}{\eta c^2},
\end{equation}
where $\eta$ is the radiative efficiency of the accretion disk, which we assume to be $\eta=0.1$ \citep[e.g.,][]{Shakura:1973aa}. Using $L_{\rm bol}^{\rm [O\;III]}$ and $L_{\rm bol}^X$, we find $\dot M_\bullet = 0.30$ and 0.11 M$_\odot$ yr$^{-1}$, respectively. We also calculate the Eddington ratio, $\lambda_{\rm Edd} = \frac{L_{\rm bol}}{L_{\rm Edd}}$, which can be thought of as a specific accretion rate. Using $L_{\rm bol}^{\rm [O\;III]}$ and $L_{\rm bol}^{X}$, we find $\lambda_{\rm Edd} =$ 0.61 and 0.22, respectively. Both of the these Eddington ratios are energetic and commensurate with what is typically found in luminous quasars \citep[e.g.][]{Aird:2012aa}. The $L_{\rm bol}^{\rm [O\;III]}$ specifically suggests near-Eddington accretion, explaining the X-ray weakness \citep{Brandt:2000aa} and making it likely this target has strong winds similar to other X-ray weak quasars \citep[e.g.,][]{Gibson:2008aa}.  These Eddington ratios also support the steep X-ray spectrum, which is indicative of a high accretion rate. 

Finally, we examine the growth rate of the supermassive black hole relative to the stellar population. We cannot separate this galaxy into bulge and disk components, so we compare the black hole mass to the total stellar mass in Figure \ref{fig:flowdiagram}. CQ 4479 lies on the $M_\bullet-M_\ast$ relation observed in local galaxies \citep{Haring:2004aa}. However, CQ 4479 still has a substantial gas reservoir remaining, so it will continue to evolve in this parameter space. We estimate in what direction this galaxy will evolve using the specific growth rates of the stellar mass and black hole mass (M$_\bullet$). For the stellar population, we measure (using the {\tt CIGALE} fit) a specific star formation rate of specific SFR, ${\rm sSFR}={\rm SFR}/M_\odot = 4.36$\,Gyr$^{-1}$. For the black hole, we measure specific black hole mass growth rate s$\dot M_\bullet = \dot M_\bullet/M_\bullet = 4.56, 11.79$\,Gyr$^{-1}$, using $L_{\rm bol}^{\rm [O\;III]}$ and $L_{\rm bol}^{\rm [O\;III]}$, respectively.
The [O {\sc iii}]-derived s$\dot M_\bullet$ is outpacing the sSFR, while the X-ray-derived s$\dot M_\bullet$ is almost exactly the same as sSFR. Assuming that the gas reservoir only lasts for 0.5\,Gyr, and the SFR and $\dot M_\bullet$ remain essentially unchanged, we calculate where CQ 4479 will end up in the $M_\bullet - M_\ast$ parameter space when its gas reservoir is depleted. Despite the different growth rates, this galaxy winds up within the scatter of the relationship currently measured for local galaxies. However, the final black hole masses differ by a factor of 3. Based on the specific growth rates, the stellar and black hole mass components of CQ 4479 are growing in lock-step, rather than the quasar phase following the starburst phase as theoretically expected \citep{Hopkins:2006aa,Kocevski:2017aa,Caplar:2018aa}. We note that with the uncertainties present in the X-ray luminosity and [O III]--L$_{Bol}$ conversion, this model remains a hypothesis to be tested by further observations.

\citet{Sun:2015aa} measured the direction of change (`flow') of  Herschel-detected broad-line AGN in the same parameter 
space. We compare CQ 4479 to their $z\leq1$ sample. Approximately 60\% of their sample have black holes that are growing proportionately faster than the stellar component at higher redshift, although most of these galaxies lie below the $M_\bullet - M_\ast$ relation. 
What is particularly interesting about CQ 4479 relative 
to the larger sample is the predicted young age of the stellar component from both the best-fit {\tt CIGALE} and {\tt SED3FIT} models. Both point toward a mass-weighted or $r$-band-weighted age of $200-500$\,Myr or shorter, similar to $\tau_{\rm dep}$. This indicates a relatively young stellar population and a galaxy that is about halfway through its star-forming lifetime. \citet{Sun:2015aa} measure an AGN duty cycle of 10\% for their sample. If CQ 4479's black hole has been accreting at roughly 0.30\,$M_\odot$ yr$^{-1}$, then it would take approximately 50\,Myr for it to build up its present BH mass, consistent with a 10\% duty cycle. However, if the galaxy is younger, or the accretion rate is lower, the duty cycle becomes much closer to unity. 


\section{Conclusions}\label{sec:conc}
The transition period between a gas-rich star-forming galaxy and its quiescent future likely includes feedback driven by an actively accreting SMBH.  To better understand how AGN are driving the evolution in their host galaxies, we conduct observations of CQ 4479 (SDSS J014040.71+001758.1) at z $\sim$ 0.405.  CQ 4479 is a rare cold quasar, host to a cold dust component and a luminous AGN.  Using newly acquired SOFIA/HAWC+ data at 89 $\micron$, we determine the following regarding CQ 4479:

\begin{itemize}
\item{ CQ 4479 is a starburst galaxy (SFR $\sim$ 95 M$_{\odot}$ yr$^{-1}$) with a cold dust component and an X-ray luminous central engine.}
\item{This object represents an early stage in AGN feedback, retaining a significant gas and dust supply with a molecular gas mass fraction of $\sim$50-70\%.}
\item{The X-ray luminosity observed is significantly lower than expected based on optical emission line or 6\,$\micron$ modeling, indicating a very recent decline in AGN energy output.}
\item{The [{\sc O iii}] emission line exhibits the potential indicator of an outflow, consistent with the recently active AGN model.}
\end{itemize}

CQ 4479 represents a rare window to observe the complex interplay between AGN and host galaxy. The detection of a cold gas supply, as well as the complex differences between X-ray bolometric luminosities, suggest a system with a potentially complex central engine that would strongly benefit from follow-up investigations.

\acknowledgements{
\linespread{1}

We thank the referee for constructive feedback that improved this work. Based in part on observations made with the NASA/DLR Stratospheric Observatory for Infrared Astronomy (SOFIA). SOFIA is jointly operated by the Universities Space Research Association, Inc. (USRA), under NASA contract NNA17BF53C, and the Deutsches SOFIA Institut (DSI) under DLR contract 50 OK 0901 to the University of Stuttgart. T.K.D.L acknowledges support from the Simons Foundation.

Based on observations made with the NASA Galaxy
Evolution Explorer. GALEX is operated for NASA
by the California Institute of Technology under NASA
contract NAS5-98034. This work is based (in part) on observations made with the Spitzer Space Telescope, which is operated by the Jet Propulsion Laboratory, California Institute of Technology under a contract
with NASA. Herschel is an ESA space observatory with
science instruments provided by European-led Principal
Investigator consortia and with important participation
from NASA. Based on data products from observations made with ESO Telescopes at the La Silla or Paranal Observatories under ESO program ID 179.A-2010.

This research made use of Astropy, a community-developed core Python package for
Astronomy \citep{Robitaille:2013aa,Price-Whelan:2018aa}. This research made use of the iPython environment \citep{Perez:2007aa} and the Python packages SciPy \citep{Virtanen:2020aa}, NumPy \citep{Walt:2011aa}, and Matplotlib \citep{Hunter:2007aa}.

The authors wish to recognize and acknowledge this work was performed on the unceded traditional land of the Kaw, Kiikaapoi, O\v{c}h\'{e}thi \v{S}ak\'{o}wi\ng{}, and Osage peoples.  We also thank Wichahpi King for their generous advisement.
}


\newcommand\invisiblesection[1]{%
  \refstepcounter{section}%
  \addcontentsline{toc}{section}{\protect\numberline{\thesection}#1}%
  \sectionmark{#1}}

\invisiblesection{Bibliography}
\bibliography{CookeKirkpatrick2020bib}

\end{document}